\documentstyle[prl,aps,twocolumn,floats]{revtex}
\begin{document}
\input epsf
\draft
\title{Inertial- and Dissipation-Range Asymptotics in Fluid Turbulence}
\author{Sujan K. Dhar, Anirban Sain, and Rahul Pandit\cite{byjnc}} 
\address{Department of Physics, Indian Institute of Science,\\
Bangalore - 560 012, India}

\date{\today}

\maketitle

\begin{abstract}
We propose and verify a wave-vector-space version of
generalized extended self similarity \cite{benzi} and broaden
its applicability to uncover intriguing, universal scaling in
the far dissipation range by computing high-order ($\leq 20\/$)
structure functions numerically for: (1) the three-dimensional,
incompressible Navier Stokes equation (with and without
hyperviscosity); and (2) the GOY shell model for turbulence.
Also, in case (2), with Taylor-microscale Reynolds numbers $4 \times
10^{4} \leq Re_{\lambda} \leq 3 \times 10^{6}\/$, we find
that the inertial-range exponents ($\zeta_{p}\/$) of
the order - $p\/$  structure functions do not approach 
their Kolmogorov value $p/3\/$ as   $Re_{\lambda}\/$ increases.
\end{abstract}



The central concern of studies of homogeneous, isotropic
turbulence is the scaling of order-$p\/$ velocity structure
functions, e.g.,  ${\cal {S}}_p(r) \equiv \langle
|{\bf{v}}_{i}({\bf{x}} + {\bf{r}}) - {\bf{v}}_{i}({\bf{x}})|^p
\rangle \/$, where $i (= 1, 2,\/$ or $3)\/$ is the Cartesian
component of the velocity ${\bf{v}}({\bf{x}})\/$ at point
${\bf{x}}\/$, and the angular brackets imply, in principle, a
spatiotemporal average. Kolmogorov (K41)
\cite{k41} predicted that, at high Reynolds numbers
$Re_{\lambda}\/$ and for the inertial range $ 20 \eta_d \lesssim r 
\ll L\/$ ($\eta_d\/$ and $L\/$ are, respectively, dissipation
and forcing scales and $\lambda\/$ is the Taylor microscale),
${\cal {S}}_p(r) \sim r^{\zeta_p}\/$ with $\zeta_p = p/3\/$.
Subsequent experimental and theoretical studies
\cite{benzi,ansel,shelev,benziol,stol,menvin,vdw,jen,pisa,lkad}
have argued for: (1) multiscaling, i.e., $\zeta_p = p/3 -
\delta \zeta_p\/$, with $\delta \zeta_p > 0\/$ but $\zeta_p\/$ a
nonlinear, monotonically increasing function of $p\/$; and (2)
extended self similarity (ESS) \cite{benziol}, in
which $\zeta_p\/$ is obtained from ${\cal {S}}_p \sim {\cal
{S}}^{\zeta_p}_3\/$, since this {\em extends} the
apparent inertial range down to $r \simeq 5 \eta_d\/$.  A
recent generalization \cite{benzi} uses ${\cal {G}}_p(r)
\equiv {\cal S}_p(r)/[{\cal S}_3(r)]^{p/3}\/$ and suggests 
that a log-log plot of ${\cal {G}}_p\/$ versus ${\cal G}_q\/$
is a straight line with slope $\rho_{p,q} = [\zeta_p - p
\zeta_3/3]/[\zeta_q - q \zeta_3/3]\/$ for the lowest resolvable
values of $r\/$.  This generalized extended self similarity
(GESS) has been tested \cite{benzi} to some extent ($p, q \leq
6\/$).

\begin{figure}[t]
\epsfxsize=19pc
\centerline{\epsfbox{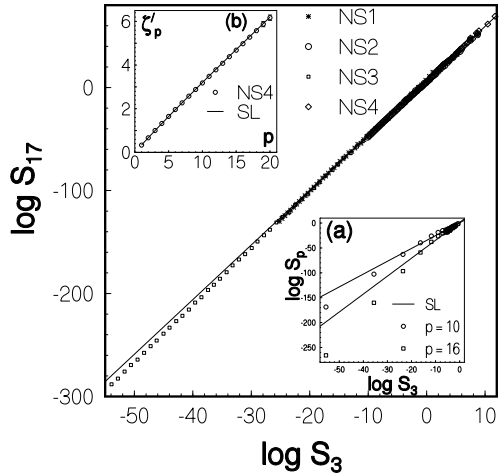}}
\caption{Log-log plots (base 10) of $S_p\/$ versus $S_3\/$ for
 $3d\/$ NS (p=17 for runs NS1-4)
 and GOY (run G1 in inset (a)) models showing our $k$-space ESS
 (Eq. 1); full lines are the SL prediction \protect\cite{shelev}. 
 Inset (b):
 $\zeta_p'\/$ (circles) from run NS4; the line is $\zeta_p' =
 2(\zeta_p + 3 p/2)/11\/$, with the $\zeta_p=\zeta_{p}^{SL}\/$.
 Note the deviation of our data points from SL lines at small
 $S_3\/$, i.e., in the dissipation range.}
\label{fig1}
\end{figure}

\begin{figure}
\epsfxsize=19pc
\centerline{\epsfbox{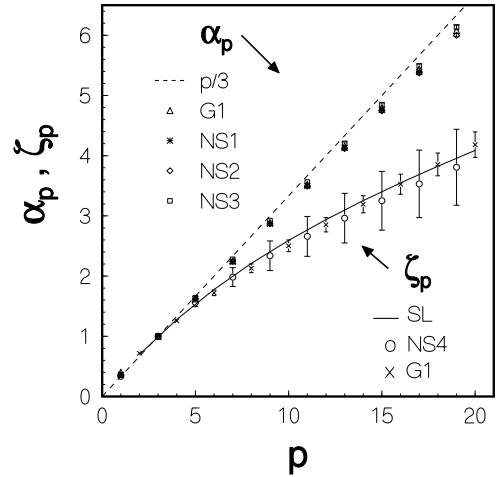}}
\caption{Inertial- and dissipation-range exponents
$\zeta_p\/$ and $\alpha_p\/$ (extracted from plots like Fig.1)
versus $p\/$ for GOY and NS runs and
their comparison with the SL formula \protect\cite{shelev} and $p/3\/$. 
We obtain
$\zeta_p\/$ from our measured $\zeta_{p}'\/$ and the formula
$\zeta_p = 11\zeta_{p}'/2 - 3p/2\/$; this amplifies the
error bars relative to Fig.1 [inset(b)]. Error bars for $\alpha_p\/$
are shown but not apparent since they are comparable to the
symbol sizes.}
\label{fig2}
\end{figure}

Here we show how GESS is modified at sufficiently small
$r\/$ by computing wave-vector-space ($k$-space)
analogs of high-order ($\leq 20\/$) structure functions for (1)
the three-dimensional, incompressible Navier Stokes equation
($3d\/$ NS), with and without hyperviscosity, and (2) the GOY
shell model for turbulence \cite{jen,pisa,lkad,goy} (where
we attain both large $Re_{\lambda}\/$ and $k \gg k_d \equiv
\eta^{-1}_d\/$). We further propose a $k$-space 
GESS \cite{benzi}, show that it holds for $L^{-1} \ll k \lesssim
1.5 k_d\/$, but then {\em crosses over to another form 
in the far dissipation range}. To study this we postulate  
$k$-space ESS (for real-space structure functions we use
the symbols ${\cal{S}}\/$ and ${\cal{G}}\/$ and for
their $k-$space analogs ({\em not} Fourier transforms) the 
symbols $S\/$ and $G\/$):
\begin{eqnarray}
S_p &\equiv& \langle |{\bf{v}}({\bf{k}})|^{p}
\rangle \approx A_{Ip} (S_3)^{\zeta'_p}, \;\;L^{-1} \ll k \lesssim 1.5
k_d, \nonumber \\
S_p &\equiv& \langle |{\bf{v}}({\bf{k}})|^{p}
\rangle \approx A_{Dp} (S_3)^{\alpha_p}, \;\;1.5 k_d \lesssim k \ll
\Lambda, 
\end{eqnarray}
where $A_{Ip}\/$ and $A_{Dp}\/$ are, respectively, nonuniversal
amplitudes for inertial and dissipation ranges 
and $\Lambda^{-1}\/$ the (molecular) length at which
hydrodynamics fails (see \cite{benziol,stol} for real-space
analogs).  Our study shows (Figs.\ref{fig1}-\ref{fig2}) that
Eq. (1) holds with two different exponents $\alpha_p\/$ and
$\zeta'_p\/$. In the GOY model $\zeta'_p = \zeta_p\/$,
but we find explicitly [inset(b), Fig.1]  that, for the $3d\/$ NS
case, $\zeta'_p = 2(\zeta_p + 3 p/2)/11\/$ (i.e., $S_p(k) \sim
k^{-(\zeta_p + 3p/2)}\/$ in the inertial range \cite{foot1}); the
difference between the two arises because of phase-space factors.
{\em Both} $\zeta_p\/$ and $\alpha_p\/$ (Fig.\ref{fig2}) seem
universal (the same for all GOY and $3d\/$ NS runs (Table
\ref{table1}) \cite{foot2}).  $\zeta_p\/$ agrees fairly with the She-Leveque
(SL)\cite{shelev} formula $\zeta_p^{SL} = p/9 + 2[1 - (2/3)^{p/3}]\/$ 
for the ranges of $p\/$ and $Re_{\lambda}\/$ in Fig.\ref{fig2}; 
and $\alpha_p\/$ is close to, but {\em systematically less} than, $p/3\/$.

\begin{figure*}[t]
\centerline{\epsfxsize=13pc \epsfbox{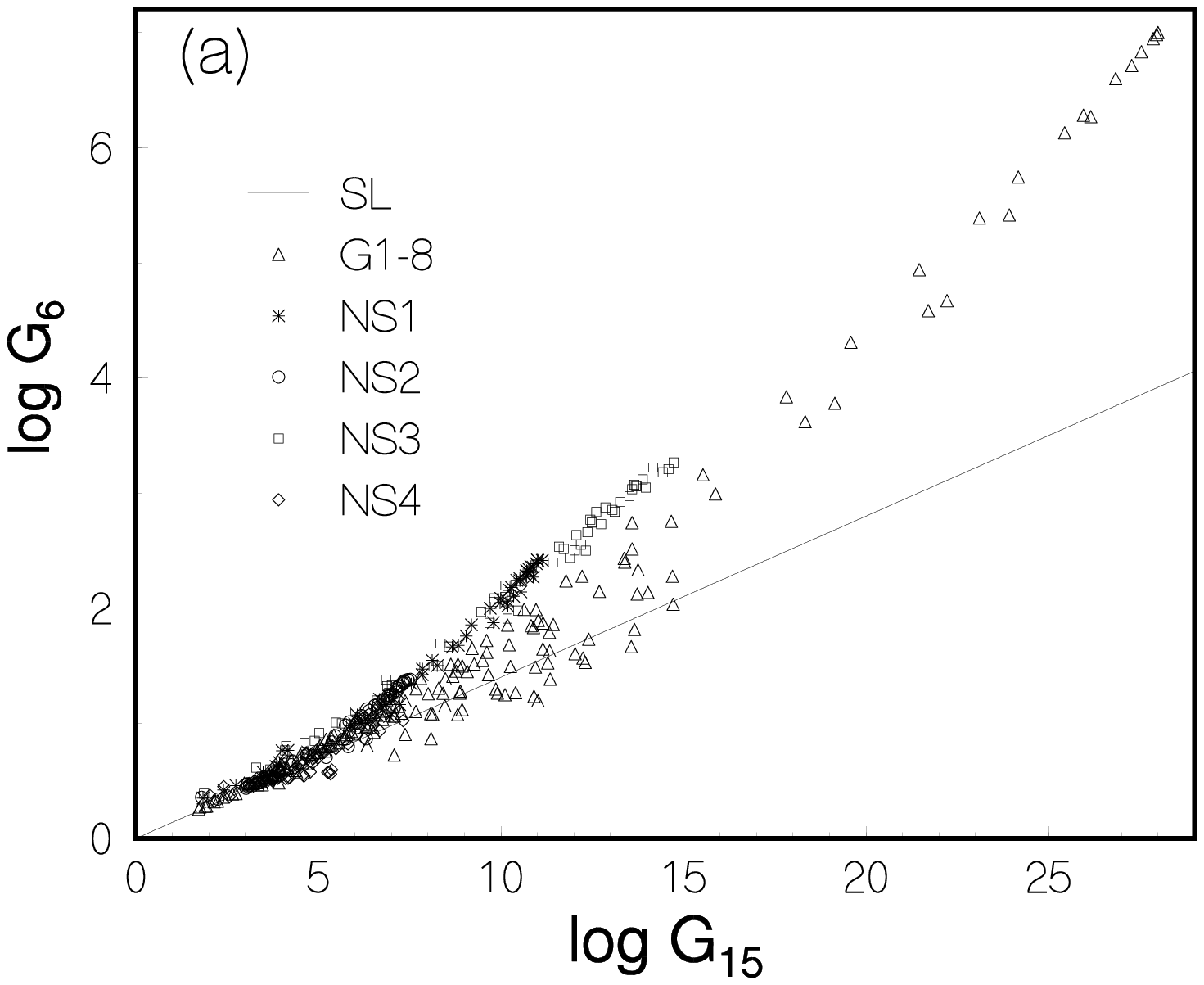} \hfill
\epsfxsize=13pc \epsfbox{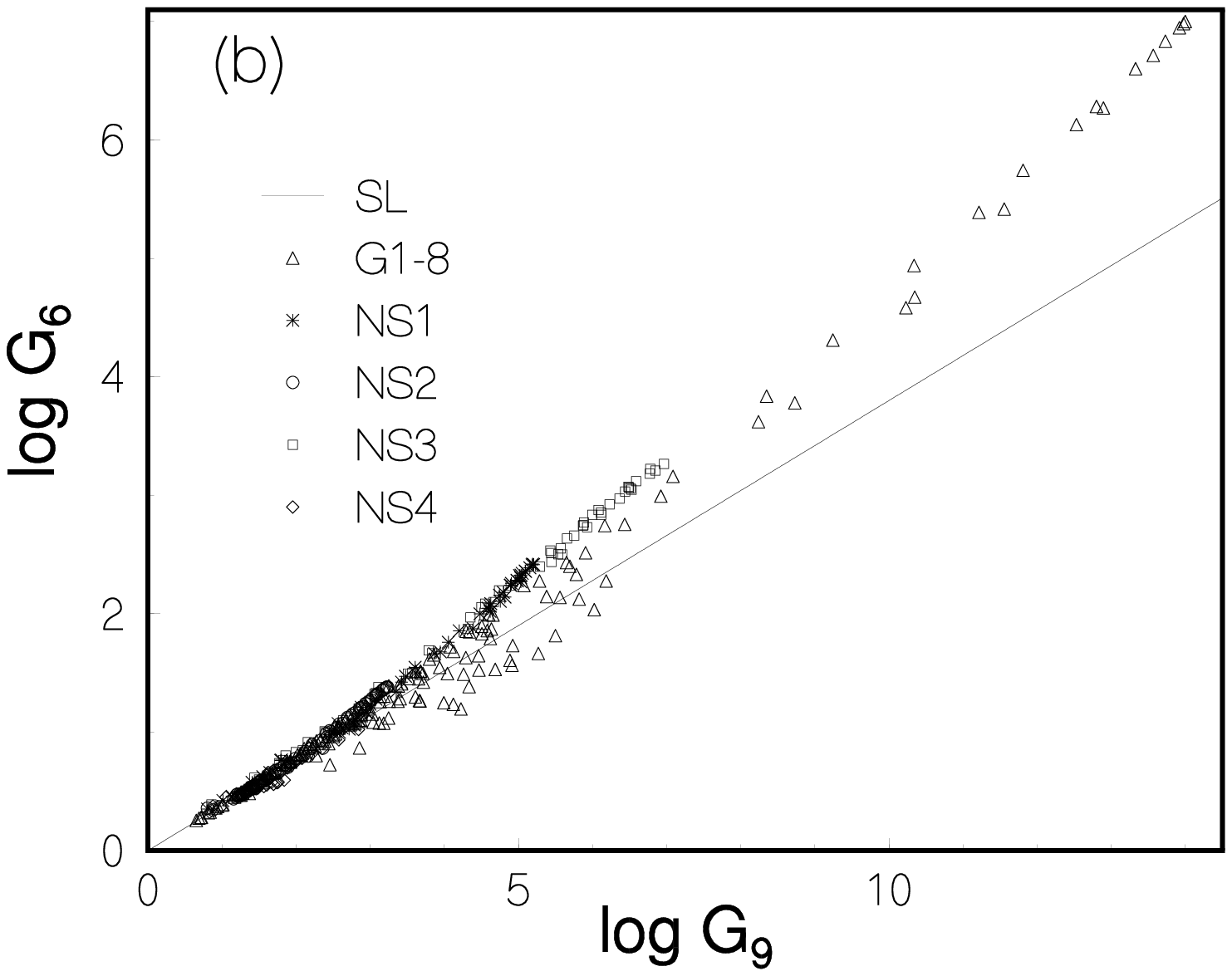} \hfill
\epsfxsize=13pc \epsfbox{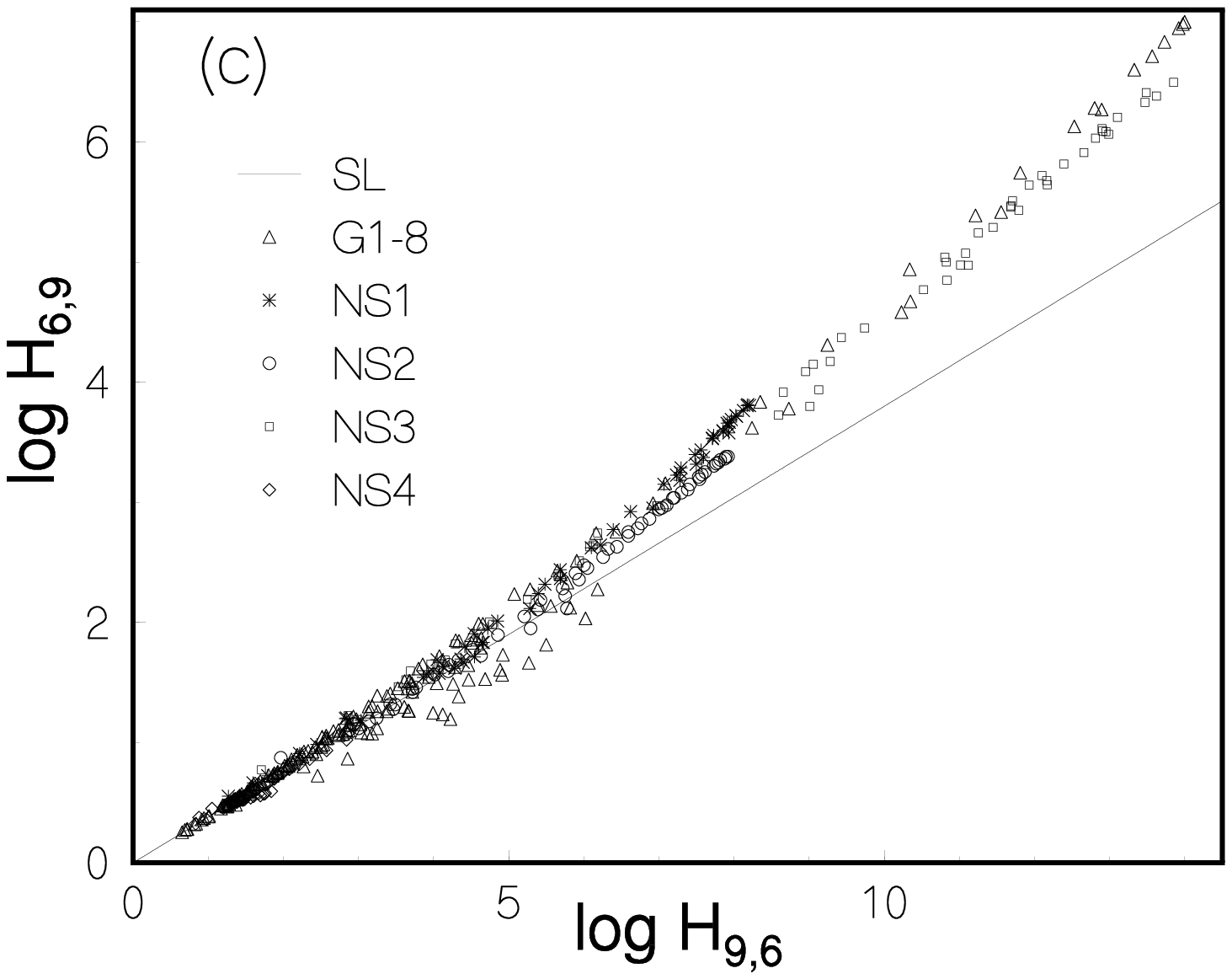}}
\caption{Log-log (base 10) plots of $G_6\/$ versus (a) $G_{15}\/$
and (b) $G_{9}\/$ illustrating our $k\/$-space GESS; (c) $H_{6,9}\/$ 
versus $H_{9,6}\/$ showing the universal inertial- to
dissipation-range crossover (see text). The line shows the SL,
inertial-range prediction.} 
\label{fig3}
\end{figure*}

The $k\/$ dependences of the inertial- and dissipation-range
asymptotic behaviors follow now from the dependence of $S_3\/$ on
$k\/$: We find 
\begin{eqnarray}
S_3 &\approx& B_I k^{-\zeta_3 - 9/2}, \;\; L^{-1} \ll k \lesssim
1.5 k_d, \\
S_3 &\approx& B_D k^{\delta} \exp(-c k/k_d), \;\; 1.5 k_d \lesssim k \ll
\Lambda,
\end{eqnarray}
where $B_I\/$ and $B_D\/$ are, respectively, nonuniversal
amplitudes (Eq. (2) holds \cite{foot1} for $3d\/$ NS; for GOY
the factor $9/2\/$ is absent). Thus, in the far
dissipation range, {\em all} $S_p \sim k^{\theta_p} \exp(- c
\alpha_p k/k_d)\/$ for $1.5 k_d \lesssim k \ll \Lambda\/$,
 with $\theta_p = \alpha_p
\delta\/$, a form not easy to verify numerically for large
$p\/$, given the rapid decay at large $k\/$, and suggested
hitherto \cite{fardis} only for $S_2\/$. In Eq. (3), $\delta, c,
k_d\/$ are not universal, but we extract the universal part of the
crossover via our $k\/$-space GESS: Define $G_{p} \equiv
S_{p}/(S_{3})^{p/3}\/$; log-log plots of $G_{p}\/$ versus
$G_{q}\/$ now yield curves (Figs. \ref{fig3}a and \ref{fig3}b)
with asymptotes which have {\em universal, but different,}
slopes in inertial and dissipation ranges. The inertial-range
asymptote has a slope $\rho(p,q)\/$ (as in real-space GESS
\cite{benzi} which 
follows from the formulae above); the resulting
$\zeta_p\/$ are in fair agreement with the SL formula \cite{shelev}.
The dissipation-range asymptote has a 
slope $\omega(p,q) \equiv [\alpha_p - p/3]/[\alpha_q - q/3]\/$
(see Eq.(1) and the definition of $G_p\/$). The slopes of these
asymptotes are universal, but the point at which the curve veers
off from the inertial-range asymptote depends on the model (GOY,
NS, etc.).  However, a simple transformation yields a {\em
universal crossover scaling function} (different for each
$(p,q)\/$ pair because of multiscaling): Define $\log(H_{pq})
\equiv D_{pq} \log(G_p)\/$ and $\log(H_{qp}) \equiv D_{qp}
\log(G_q)\/$; the scale factors $D_{pq} = D_{qp}\/$ are {\em
nonuniversal}, but 
plots of $\log(H_{pq})\/$ versus $\log(H_{qp})\/$ show data from
{\em all} GOY and $3d\/$ NS runs  collapsing onto {\em one
universal curve} within our error bars (Fig.\ref{fig3}c for $p = 6\/$
and $q = 9\/$) for {\em all} $k\/$ and $Re_{\lambda}\/$.
Both ESS (Fig.\ref{fig1}) and GESS (Fig.\ref{fig3}) remove the exponential
{\em controlling factor} \cite{benors} from the {\em leading
asymptotic behavior} of $S_p\/$ in the far dissipation range and
expose the remaining power-law dependence on $k\/$. 
Also, it is easy to see analytically that GESS plots (Fig.\ref{fig3}) amplify
slope differences between inertial- and dissipation-range
asymptotes relative to ESS plots (Fig.\ref{fig1}).

\begin{figure}[t]
\epsfxsize=19pc 
\centerline{\epsfbox{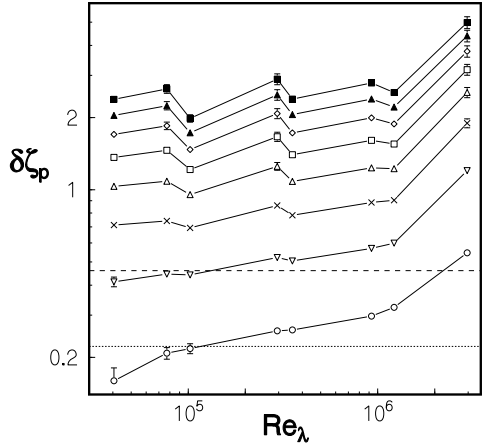}} 
\caption{Log-log plot(base 10) of $\delta \zeta_p\/$ versus
the Taylor-microscale Reynolds number $Re_{\lambda}\/$ for our GOY runs (G1-8)
with $p = 6,8 \ldots, 20\/$ (from bottom to top). The dotted
($p = 6\/$) and dashed ($p = 8\/$) lines show the SL results 
\protect\cite{shelev}. Error bars are shown
 but are often smaller than the symbol sizes.} 
\label{fig4}
\end{figure}

How robust is the fair agreement of $\zeta_p\/$ (Fig.\ref{fig2}) with
the SL formula?  Some studies \cite{katsu,proc1,proc2} suggest
that, as $Re_{\lambda} \rightarrow \infty\/$, $\delta \zeta_p\/
\equiv (p/3 - \zeta_p) \rightarrow 0\/$. Numerical solutions of
the $3 d\/$ NS equation can at best achieve \cite{menvin,lkad,cdkw}
$Re_{\lambda} \lesssim 220\/$, too small, by far, to resolve
this issue, so we address it for the GOY model, by studying the
range $4 \times 10^{4} \lesssim Re_{\lambda} \lesssim 3 \times
10^{6}\/$. We find (Fig.\ref{fig4}) that $\delta \zeta_{p}$ does not
vanish with increasing $Re_{\lambda}\/$; if anything,
it rises marginally \cite{foot3}. Systematic experimental studies at high
$Re_{\lambda}\/$ are perhaps the best way to check if the
trends of Fig.\ref{fig4}  obtain in the $3d\/$ NS case.

We remark that, if we {\em assume} the hierarchy 
$[G_{p+1}/G_p] = [G_{p}/G_{p-1}]^{\gamma}
\times [\lim_{p \rightarrow \infty} G_{p+1}/G_p]^{1 - \gamma}\/$ with
$\gamma^3 = 2/3\/$ (whose real-space analog is equivalent
\cite{benzi} to the SL moment hierarchy for the energy dissipation
\cite{shelev}) and use \cite{foot4} $G_p(k)
\approx C_p k^{\beta_p}\/$, we get a difference equation for
$\beta_p\/$ {\em identical} to the SL one (our $\beta_p\/$ is 
their $-\tau_{p/3}\/$).  This, when solved
with the boundary conditions $\beta_0 = \beta_3 = 0\/$ and
$\lim_{p\rightarrow\infty} (\beta_{p+1} - \beta_p) = 2/9\/$, 
yields the SL formula (via $\zeta_p = -\beta_p + p\zeta_3/3\/$).
However, our GESS yields $[G_{p+1}/G_p] \approx C'_{p}
[G_{p}/G_{p-1}]^{\Upsilon_p}\/$ with $\Upsilon_p = (\zeta_{p+1}
- \zeta_p - 1/3)/(\zeta_{p} - \zeta_{p-1} - 1/3)\/$.
Superficially, this might seem to violate the hierarchy assumed above,
but it turns out to be consistent with our GESS form, if $\Upsilon_p
=\gamma - 2(1-\gamma)/[9(\zeta_p-\zeta_{p-1}-\zeta_3)]\/$, which
is precisely the SL difference equation. Of course, our GESS form 
can hold with $\zeta_p\neq\zeta_p^{SL}\/$; Fig. \ref{fig2}
shows the quality of agreement between our measured $\zeta_p\/$ 
and $\zeta_p^{SL}\/$. 

\widetext
\begin{table*}[t]
\caption{Parameters $\nu\/$ (viscosity), $\nu_H\/$ (hyperviscosity), 
$Re_{\lambda}\/$ (Taylor-microscale Reynolds number), $\tau_e\/$ (box-size
eddy-turnover time), $\tau_{av}\/$ (averaging time), $\tau_t\/$ (transient
time) and $k_d\/$ (dissipation-scale wavenumber) for our $3d\/$ NS runs NS1-4
($k_{max} = 64\/$) and GOY-model runs G1-8 ($k_{max} = 2^{22}
k_0\/$). The step size($\delta t\/$) used is 0.02 for NS1-4, $10^{-4}\/$ for
G1-4, and $2\cdot10^{-5}\/$ for G5-8.} 
\label{table1}
\begin{tabular}{|l|c|c|c|c|c|c|c|}
\hline
Run & $\nu$ & $\nu_H$ & $Re_{\lambda}$ & 
$\tau_e/\delta t\/$ & $\tau_t/\tau_e\/$ & $\tau_{av}/\tau_e\/$ &
$k_{max}/k_d\/$ \\ \hline 
NS1 & $5\cdot 10^{-4}$ & $0$ & $\simeq 3.5$ & 
$\simeq 3 \cdot  10^4\/$ & $\simeq 1 \/$ & $2 \/$ &  $\simeq 4\/$ \\
NS2 & $2 \cdot  10^{-4}$ & $0$ & $\simeq 8\/$ & 
$\simeq 3 \cdot  10^4\/$ & $\simeq 1 \/$ & $\simeq 2.5 \/$ &  $\simeq 2.3\/$\\
NS3 & $5 \cdot  10^{-4}$ & $5 \cdot  10^{-6}$ & $\simeq 3.5$ &
$\simeq 3 \cdot  10^4\/$ & $\simeq 1\/$ & $\simeq 1
\/$ & $\simeq 6.5\/$\\ 
NS4 & $5 \cdot  10^{-4}$ & $10^{-6}$ & $\simeq 22$ &
$\simeq 3 \cdot  10^3\/$ & $\simeq 10 \/$ & $\simeq 7
\/$ & $\simeq 2\/$\\ 
G1-4 & $5\cdot 10^{-6} - 10^{-7}$ & $0$ & $4 \cdot  10^4 - 3 \cdot 
10^5$ &  $\simeq(1.5-2.0)10^4\/$ & $\simeq 500\/$ &
$\simeq 2500\/$ & $\simeq 2^5 - 2^3\/$\\ 
G5-8 & $5 \cdot  10^{-8} - 10^{-9}$ & $0$ & $3.5 \cdot  10^5 - 3 \cdot 
10^6$ &  $\simeq (0.7-1)10^5\/$ & $\simeq 500 \/$
& $\simeq 2500\/$ & $\simeq 2^3 - 1\/$\\ 
\hline
\end{tabular}
\end{table*}

\narrowtext

We use a pseudospectral method \cite{menvin} for our 
numerical solution of the incompressible $3d\/$ NS
equation. We force the first two $k\/$-shells, use a
box with side $L_B = \pi\/$ and  $64^3\/$ modes. Our
dissipation term $-(\nu + \nu_H k^2) k^2\/$ allows for both
viscosity $\nu\/$ and hyperviscosity $\nu_H\/$. For time
integration we use an Adams-Bashforth scheme (step size $\delta
t\/$) \cite{menvin}.  Parameters for our $3d\/$ NS runs NS1-4 are
given in Table \ref{table1}, where $\tau_e \equiv L_B/v_{rms}\/$
is the box-size eddy-turnover time and $\tau_{av}\/$ the
 averaging time, 
after initial transients have decayed over a period $\tau_t\/$.
We use $Re_{\lambda}\/ \equiv v_{rms} \lambda / \nu\/$, where $
\lambda = [\int_0^{\infty} E(k) dk/\int_0^{\infty} k^2 E(k)
dk]^{1/2}\/$, $v_{rms} = [(2/3L_B^3) \int_0^{\infty} E(k) dk]^{1/2}\/$
and $E(k) \sim S_2(k) k^2\/$. All $S_p(k)\/$ are averaged over
shells of radius $k\/$.  
Care must be exercised in choosing $\delta t\/$
and the forcing amplitude, otherwise there is a slow, but
systematic, stretching of the data points along the asymptotes
in Figs. \ref{fig1} and \ref{fig3} with
increasing $\tau_{av}\/$ (over the time scales of our low-$Re_{\lambda}\/$ 
NS runs).  Fortunately, this hardly affects our exponents:
any attendant systematic errors in Fig. \ref{fig2}
are certainly less than the random errors indicated.
Also, the agreement between our GOY and NS runs confirms our
results. Our GOY-model data are, of course, of much better quality.
Here Fourier components of the velocity are labeled by a
discrete set of wave vectors $k_n = k_0 q^n\/$. The dynamical
variables are the {\em complex, scalar} velocities $v_n\/$ for
each shell $n\/$; $v_n\/$ is affected directly only by the
velocities in nearest and next-nearest shells. In spite of its
simplicity, this model yields scaling properties
\cite{jen,pisa,lkad,goy} akin to experimental ones. The
GOY-model equations are: 
\begin{equation}
\frac{d}{dt}v_n = i C_n - \nu k_n^2 v_n + f_n,
\end{equation}
where $\nu\/$ is the kinematic viscosity, $f_n\/$ the external
force on shell $n\/$, $C_n = ( a k_n v_{n+1} v_{n+2} + b k_{n-1}
v_{n-1} v_{n+1} + c k_{n-2} v_{n-1} v_{n-2})^{\ast}\/$, and $a,
b,\/$ and $c\/$ can be fixed upto a constant by demanding
\cite{lkad}, for $\nu , f_n = 0\/$, that: $v_n 
\sim k_n^{-1/3}\/$ be a stationary solution of Eq.(4); and 
the GOY-model kinetic energy and helicity be
conserved. We adopt the conventional parameters \cite{pisa,lkad}
$k_0 = 2^{-4}, \; q = 2, \; a = 1, \; b = c = -1/2\/$ and use
$f_n = 5\cdot 10^{-3}(1+i) \delta_{n,1},\/$ i.e., we force the
first shell \cite{foot5}. The GOY-model
structure functions are $S_{n,p} \equiv \langle |v_n|^p \rangle
\sim k_n^{-\zeta_p}\/$ \cite{jen,pisa,lkad}; reliable values of
$\zeta_p\/$ obtain \cite{lkad} if we use 
$\Sigma_{n,p} = \langle |\Im [v_n v_{n+1} v_{n+2} + v_{n-1} v_n
v_{n+1}/4]|^{p/3} \rangle\/$ since this removes an underlying
$3-\/$cycle. We have used $\Sigma_{n,p}\/$ to obtain
Fig.\ref{fig4} \cite{foot6}, but
$S_{n,p}\/$ in Figs.\ref{fig1}-\ref{fig3} for consistency with $3d\/$ NS.
We use an Adams-Bashforth scheme \cite{pisa}
(step size $\delta t\/$) to integrate Eq. (4).  The average
of the time scale associated with the smallest wavenumber, $(|v_1|
k_1)^{-1}\/$, gives the ``box-size'' eddy turnover time.  Table \ref{table1}
lists other parameters for our 8 GOY-model 
runs G1-8, for which we use (cf., \cite{pisa}) $E(k) =
S_{n,2}/k_n\/$, $\lambda = 2 \pi/k_0[\sum_n S_{n,2}/\sum_n k_n^2
S_{n,2}]^{1/2}\/$, and $v_{rms} = [k_0 \sum_n
S_{n,2}/\pi]^{1/2}\/$. This yields $Re_{\lambda} \sim
\nu^{-0.5}\/$, as expected \cite{lohprl} at large
$Re_{\lambda}\/$.

Experimental evidence for the slope change in the dissipation
range in real-space analogs of Fig.\ref{fig1} was given by Stolovitzky
and Sreenivasan \cite{stol}, who postulated ${\cal{S}}_p \sim
{\cal{S}}_3^{\alpha'_p}\/$ in the dissipation range and
suggested $\alpha'_p \simeq (\zeta_{3p/2} + p/2)/(\zeta_{9/2} +
3/2)\/$. We have not been able to obtain a simple relation
between our $\alpha_p\/$ and their $\alpha'_p\/$ (unlike
\cite{foot1} that between $\zeta_p\/$ and $\zeta'_p\/$)
since $S_p\/$ does not have a power-law dependence on $k\/$ in
the dissipation range. It would be very interesting to extend
such experimental studies to test the {\em universality} of
dissipation-range asymptotics (e.g., in different flows) and the
crossover suggested here. The universal multiscaling in the
dissipation range that we have elucidated is a manifestation of
strongly intermittent 
(multifractal) dissipation which is believed to occur
\cite{fardis} even at low $Re_{\lambda}\/$.  We believe that this
multiscaling should extend far enough into the dissipation range
before corrections set in because of the breakdown of (a) the
incompressibility assumption (at large Mach numbers) and/or (b)
hydrodynamics (at molecular length scales). Preliminary studies
\cite{abasu} yield similar phenomena in MHD turbulence.

We thank S. Ramaswamy for discussions, CSIR and BRNS (India) for
support, and SERC (IISc, Bangalore) for computational resources.


\begin{references}
\bibitem{foot3} The increase in $\delta \zeta_p\/$ for run G8
did not go away on reducing $\delta t\/$ to $4\times10^{-6}\/$, with
$\tau_t \simeq 500 \tau_e\/$ and $\tau_{av} \simeq 2500 \tau_e\/$.
\bibitem{foot4} In the dissipation range, $G_p \sim k^{\beta'_p}
\exp(-c_p'k/k_d)\/$, so there is no SL analog for our
dissipation-range exponents.
\end{references}
\end{document}